\documentclass[aps,pra, reprint]{revtex4-1}

\usepackage{amsmath,amssymb}
\usepackage[table]{xcolor}
\usepackage{hyperref,graphicx}
\usepackage{multirow}

\newcommand{\ket}[1]{|#1\rangle}

\newcommand{\figref}[1]{Fig.~\ref{#1}}

\newcommand{\perm}{\mathrm{perm}\,}
\newcommand{\qed}{~$\square$}

\newtheorem{prop}{Proposition}
\newtheorem{corol}{Corollary}[prop]

\begin{document}

\title{Suppression laws for multi-particle interference in Sylvester interferometers}

\author{Andrea Crespi}
\email{andrea.crespi@polimi.it}
\affiliation{Istituto di Fotonica e Nanotecnologie - Consiglio Nazionale delle Ricerche,\\ p.za Leonardo da Vinci 32, 20133-Milano, Italy}
\begin{abstract}
Quantum interference of correlated particles is a fundamental quantum phenomenon which carries signatures of the statistics properties of the particles, such as bunching or anti-bunching. In presence of particular symmetries, interference effects take place with high visibility, one of the simplest cases being the suppression of coincident detection in the Hong-Ou-Mandel effect. Tichy \textit{et al.} recently demonstrated a simple sufficient criterion for the suppression of output events in the more general case of Fourier multi-port beam splitters. Here we study the case in which $2^q$ particles (either bosonic or fermionic) are injected simultaneously in different ports of a Sylvester interferometer with $2^p \geq 2^q$ modes. In particular, we prove a necessary and sufficient criterion for a significant fraction of output states to be suppressed, for specific input configurations. This may find application in assessing the indistinguishability of multiple single photon sources and in the validation of boson sampling machines.
\end{abstract}

\maketitle

\section{Introduction} \label{sec:intro}

Multi-particle quantum interference arises when several indistinguishable particles have non-vanishing probability amplitude of being found at the same site or spatial coordinate. The algebraic sum of all these probability amplitudes may lead to strong enhancement (constructive interference) or suppression (destructive interference) of the detection probability of the different possible collective states. In that, it is a pure and typical quantum phenomenon, which is worth being investigated both from a fundamental perspective and for its quantum information implications.

Qualitatively different behaviours may be observed in general, depending on the bosonic or fermionic nature of the particles. The anti-symmetrization requirements \cite{girardeau1965psm} for fermionic wavefunctions lead to vanishing probability of finding more than one particle on the same site (Pauli principle). Bosons, on the contrary, show a remarked tendency to bunch together, with increased probability of coalesce on the same site \cite{arkhipov2011bbp,spagnolo2013grb} or to cluster in nearby sites (bosonic clouding \cite{carolan2014evq}).

However, when particles evolve following Hamiltonians with specific symmetries, particular fine-grained distributions can be observed with enhanced interference peaks and dips. The simplest case is the Hong-Ou-Mandel (HOM) effect when two particles impinge on distinct ports of a balanced beam splitter: quantum interference suppresses the coincident output (one particle per each output) in the case of bosons and the single-port output (both particles in either output) in case of fermions. In the multi-particle case, a sort of generalised HOM effect occurs for symmetric multiport beam splitters \cite{ou1999ofp,campos2000tph,tichy2010ztl,spagnolo2013tpb}. In particular, Tichy et al. \cite{tichy2010ztl} showed that for a particular class of multiports, namely Bell or Fourier multiports, and input states with cyclic symmetry, a full suppression of most of the output combinations is observed; a simple analytical law gives a sufficient criterion for such suppression.

From a computational point of view, calculating the output distribution of a number of indistinguishable bosons is a mathematically \textit{hard} problem, in that it cannot be performed efficiently on conventional (classical) computers. In fact, it relies on the calculation of permanents of matrices, for which an efficient classical algorithm is lacking. The realisation of such difficulty has led to the proposal \cite{aaronson2011ccl} of boson sampling devices as experimentally accessible platforms that could perform some task hard-to-simulate with classical resources. The specialised task of such quantum devices is to physically implement and sample the distribution of $n$ bosons undergoing a certain unitary evolution. The computational difficulty of a classical simulation of such process (i.e. a classical sampling of such distribution) increases exponentially with $n$, rapidly becoming infeasible. First proof of principle experiments with photons have been reported very recently \cite{spring2013bsp,broome2013pbs,tillmann2013ebs,crespi2013imi,carolan2014evq,spagnolo2014evp}; while not having demonstrated yet a true quantum supremacy, these experiments have pointed out that such a demonstration may not be so far.

A future many-modes boson sampling experiment will likely require the implementation of an arbitrary unitary matrix, through a possibly reconfigurable \cite{shadbolt2012gmm} linear interferometer \cite{reck1994era,crespi2013imi}. If the output is hard to predict classically, it may be not trivial also the certification of the correct operation of such device: in fact, several solutions to this problem have been debated \cite{aaronson2013bfu,carolan2014evq,spagnolo2014evp,tichy2014sea,walschaers2014sbb}. The use of particular symmetric unitaries that show rich but easily predictable multiphoton distributions has been also proposed \cite{tichy2014sea} as a convenient way to assess both the performance of a multiphoton source, as well as of the reconfigurable device itself. In the same way the two-photon HOM effect on balanced beam-splitters has long been used as a diagnostic instrument for measuring two-photon indistinguishability, suppression laws for multi-port interferometers could provide a suitable means to simultaneously test the quality of a multi-photon source and of a multi-mode reconfigurable device for boson sampling experiments. Of course, these could be adopted to assess the performance of a multi-photon source also outside of the boson sampling context.

It has to be noted that the existence of a sharp suppression law directly comes from the symmetry characteristics of the matrix. While it has been conjectured that other suppression laws could exist for other class of symmetric unitaries, only the class of Fourier matrices has been investigated extensively up to now\cite{lim2005gho,tichy2010ztl, tichy2012mpi}.

In this work we address the study of interferometers implementing $m=2^p$-modes Sylvester matrices and prove a necessary and sufficient criterion for the suppression of most output combinations, for certain input states of $n=2^q \leq m$ particles, either bosonic or fermionic. We further discuss possible applications for assessing the indistinguishability of multiple single-photon sources.

In Section \ref{sec:prel} we recall some basic concepts about the evolution of multi-particle Fock states through linear unitary processes and about the definition of Sylvester matrices. In Section \ref{sec:twoPhot} we give a comprehensive characterization of the output distributions in the two-particle case, while in Sections~\ref{sec:multiPhot} and \ref{sec:fermions} we prove the output suppression criteria for the cases of multiple bosons and multiple fermions respectively. These sections are mainly organised as a list of propositions regarding mathematical properties of certain matrices and their consequences on the calculation of the multi-particle output distributions.
Finally, we discuss in Section \ref{sec:disc} the consequences and possible applications of the suppression criteria proved in the preceding sections, with particular regard to experiments with photons.

\section{Preliminary concepts} \label{sec:prel}

\subsection{Multi-particle interference in linear interferometers}

A generic Fock state $\ket{T}$ of $n$ particles on $m$ modes can be written as $\ket{T} = \left( \prod^n_{i=1} a^{\dagger}_{t_i} \right) \ket{0}$ where $a^{\dagger}_{t_i}$ is the creation operator on the mode $t_i$. Such state can be identified by the $n$-element vector $\vec{t} = \left(t_1, t_2, \ldots, t_n \right)$, with $1 \leq t_i \leq m$.  Since different orderings of the particles in the same modes are not distinguishable, we will consider only the cases $t_1 \leq t_2 \leq \ldots \leq t_n$. 

An $m$-mode lossless linear evolution can be described by a $m \times m$ unitary transformation $U$ on the space of creation operators. The probability amplitude associated with an input $\vec{g} = \left(g_1, g_2, \ldots, g_n \right)$ and output $\vec{h} = \left(h_1, h_2, \ldots, h_n\right)$ is given by
\begin{equation} \label{eq:permanent}
p_{\mathrm{bos}} = \frac{\perm S_{\vec{g},\vec{h}}}{\sqrt{\mu_1! \mu_2! \ldots \mu_m! \nu_1! \nu_2! \ldots 	\nu_m!}}
\end{equation}
in the case of bosons and by
\begin{equation} \label{eq:determinant}
p_{\mathrm{fer}} = \frac{\det S_{\vec{g},\vec{h}}}{\sqrt{\mu_1! \mu_2! \ldots \mu_m! \nu_1! \nu_2! \ldots 	\nu_m!}}
\end{equation}
where $\mu_i$ and $\nu_i$ are the number of particles present in mode $i$ in the $g$ and $h$ states respectively, $S$ is the scattering matrix with elements $S_{i,j} = U_{h_i,g_j}$ and $\perm A$ denotes the permanent of a matrix $A$.
It may be useful to recall also the definition of permanent for a matrix $A$:
\begin{equation} \label{eq:permDef}
\perm  A = \sum_\sigma \prod^n_{i=1} a_{i,\sigma(i)}
\end{equation}
where $a_{i,j}$ is an element of $A$, $\sigma$ is a permutation of $\lbrace 1, \ldots, n \rbrace$ and thus the sum in the expression is performed over all the possible permutations.

\subsection{Hadamard, Fourier and Sylvester matrices}

A complex Hadamard matrix is defined as an orthogonal matrix of complex numbers, in which all the elements have unitary modulus. A well-known sub-class of such matrices is that of Fourier matrices, the elements of a $m \times m$ Fourier matrix $F(m)$ being defined as follows:
\begin{equation}
F_{j,k} = e^{2\pi \iota (j-1) (k-1) / m }
\end{equation} 
where  $\iota = \sqrt{-1}$ is the imaginary unit. As already mentioned, multi-particle interference has been largely studied in the literature\cite{lim2005gho,tichy2010ztl,tichy2012mpi} for interferometers implementing the normalized (unitary) version of such matrices $U_{m} = \frac{1}{\sqrt{m}} F(m)$.

Real Hadamard matrices, simply referred to as Hadamard matrices in the following, are orthogonal matrices with all elements equal to $\pm 1$. Sylvester matrices are a particular class of real Hadamard matrices, having size $m = 2^p$, that can be built recursively from the following formula:
\begin{equation} \label{eq:sylvester}
H(2^p) = \begin{bmatrix} H(2^{p-1}) &  H(2^{p-1}) \\ H(2^{p-1}) & - H(2^{p-1}) \end{bmatrix}
\end{equation}
being $H(2^0) = H(1) = [1]$.
From this construction one can derive an analytic expression for the $(i,j)$ element of the matrix
\begin{equation} \label{eq:sylvesterElement}
\left[ H(2^p) \right]_{i,j} = (-1)^{i_B \odot j_B}
\end{equation}
where $i_B$ and $j_B$ are the binary representations of $i$ and $j$, enumerating the rows and columns starting from 0, and $\odot$ is the bitwise dot-product.

In the following we will refer to devices implementing a unitary matrix of the kind:
\begin{equation}
U_{m} = \frac{1}{\sqrt{m}} H(m)
\end{equation}
with $m=2^p$ as \textit{Sylvester interferometers}.

A general expression for the permanent of Sylvester (and more in general, Hadamard) matrices is not known: while $\perm H(2) = 0$, it has been conjectured that for all the other orders Hadamard matrices have non-vanishing permanents \cite{cheon2005ums}.

Note that the usual balanced beam-splitter operator is just $ \tfrac{1}{\sqrt{2}} H(2)$. In this case, for an input state with one photon per mode, the well known Hong-Ou-Mandel effect is observed, which consists in the suppression of the output state with one photon per mode. In fact, according to \eqref{eq:permanent}, for such output contribution, the probability amplitude is proportional to $\perm  H(2) = 0$. 

\section{Two particles} \label{sec:twoPhot}

Even though some of the results of this section could be retrieved by applying  the more general results of Sections~\ref{sec:multiPhot} and \ref{sec:fermions}, the two-particle case allows for a more comprehensive description and shows some specific feature, which make it worth addressing it separately.

\begin{prop} \label{twoPhotonFirstTwoRows}
If two \emph{bosons} are injected in the first two modes of an interferometer described by $U_{m} =  \frac{1}{\sqrt{m}} H(m) = \frac{1}{\sqrt{2^p}} H(2^p)$, the probability amplitude $p_{i,j}$ of an output state with one particle on mode $i$ and one particle on mode $j$ follows the rule:
\begin{align*}
&|p_{i,j}| = \frac{1}{2^{p-1}} &\mathrm{if} \quad i\mod 2 = j\mod 2 ,\; i \neq j\\
&|p_{i,j}| = \frac{1}{2^{p-1/2}} &\mathrm{if} \quad i = j\\
&p_{i,j} = 0 &\mathrm{else }
\end{align*}
\end{prop}

The scattering matrices $S$ in \eqref{eq:permanent}, for such an input state, are all sub-matrices of the first two columns of $U_m$. 
Reminding \eqref{eq:sylvester}, one can easily observe that such columns are just (properly normalized) repetitions of $H(2)$ and retrieve the matrices $S$, as a function of $i$ and $j$, as follows:
\begin{itemize}
\item $i \mod 2 = 1 \quad \mathrm{and} \quad j \mod 2 = 0\\ S = \tfrac{1}{\sqrt{2^p}} H(2) = \tfrac{1}{\sqrt{2^p}} \begin{bmatrix} 1&1\\ 1&-1 \end{bmatrix} \, \Rightarrow \, \perm S = 0$ 
\item $i \mod 2 = 0 \quad \mathrm{and} \quad j \mod 2 = 1\\  S =  \tfrac{1}{\sqrt{2^p}} \begin{bmatrix} 1&-1\\ 1&1 \end{bmatrix} \, \Rightarrow \, \perm S = 0$ 
\item  $i \mod 2 = j \mod 2$\\ $S$ consists of two identical rows with elements $\pm \tfrac{1}{\sqrt{2^p}}$: simple calculations show that in this case $| \perm S | =  \tfrac{1}{2^{p-1}}$
\end{itemize}
Thus, if and only if $i \mod 2 \neq j \mod 2$ the permanent of the scattering matrix vanishes, giving $p_{i,j} = 0$, while the other cases are proved by applying \eqref{eq:permanent}.\qed

\begin{corol} \label{twoPhotonFraction}
In the case of Prop.~\ref{twoPhotonFirstTwoRows}, the fraction of suppressed states is $\frac{N^\mathrm{bos}_\mathrm{supp}}{N_\mathrm{states}}=\frac{1}{2}\frac{m}{m+1}$.
\end{corol}

The possible two-bosons output states are identified by all the couples $(i,j)$ with $1 \leq i \leq j \leq m$ (we consider only $i \leq j$ because $(i,j)$ is the same state as $(j,i)$). The number of such states is $\frac{1}{2} m \left( m+1 \right) $. In a chessboard with $m \times m$ squares, alternately black and white, these can be seen as all the squares above the main diagonal or included in it. If we colour the squares in such a way that the main diagonal is black, the condition $i \mod 2 \neq j \mod 2$ (with $i \leq j$) indicates all the white squares comprised in the region above it, which are actually half of the total number of white squares. Thus, the number of suppressed states is $\frac{1}{4} m^2$, giving the result.\qed

\begin{prop} \label{twoFermionsFirstTwoRows}
If two \emph{fermions} are injected in the first two modes of an interferometer described by $U_{m} =  \frac{1}{\sqrt{m}} H(m) = \frac{1}{\sqrt{2^p}} H(2^p)$, the probability amplitude $p_{i,j}$ of an output contribution with one particle on mode $i$ and one particle on mode $j$ follow the rule:
\begin{align*}
&|p_{i,j}| = \frac{1}{2^{p-1}} &\mathrm{if} \quad i\mod 2 \neq j\mod 2\\
&p_{i,j} = 0 &\mathrm{else }
\end{align*}
\end{prop}

The scattering matrices to be considered for calculating the probability amplitudes are just the same of Prop.~\ref{twoPhotonFirstTwoRows}, but the determinant (Eq.~\eqref{eq:determinant}) instead of the permanent has to be calculated here. Thus, when $i\mod 2 \neq j\mod 2$ the scattering matrix is composed of two identical rows and the determinant vanishes. In all the other cases (see the expression of $S$ in the proof of Prop.~\ref{twoPhotonFirstTwoRows}) the determinant is equal to $\frac{1}{2^{p-1}}$. Application of ~\eqref{eq:determinant} then gives the probability amplitudes.\qed

\begin{corol} \label{twoFermionsFraction}
In the case of Prop.~\ref{twoFermionsFirstTwoRows} the fraction of suppressed states is $\frac{N^\mathrm{fer}_\mathrm{supp}}{N_\mathrm{states}}=\frac{1}{2}\frac{m+2}{m+1}$.
\end{corol}

Comparing Prop.~\ref{twoPhotonFirstTwoRows} with Prop.~\ref{twoFermionsFirstTwoRows}, bosons and fermions show a dichotomic behaviour, in that an output combination is suppressed for two bosons if and only if is allowed for two fermions and vice versa. Thus the fraction of suppressed states (over all the possible two-particle states) for two fermions is $\frac{N^\mathrm{fer}_\mathrm{supp}}{N_\mathrm{states}} = 1 -\frac{N^\mathrm{bos}_\mathrm{supp}}{N_\mathrm{states}}= 1 -\frac{1}{2}\frac{m}{m+1} = \frac{1}{2}\frac{m+2}{m+1}$.\qed

Note that this fraction actually includes some states (the states with two particles on the same port) that are indeed suppressed by virtue of the Pauli principle and not by specific features of the Sylvester matrix.

\begin{prop} \label{twoPhotonArbitraryInputs}
For $U_{2^p}=\frac{1}{\sqrt{2^p}} H(2^p)$ and a two-particle input on an arbitrary couple of different modes, the number of suppressed states is the same as that given in Corol.~\ref{twoPhotonFraction} and Corol.~\ref{twoFermionsFraction} for bosons and fermions respectively.
\end{prop}

For an arbitrary input state, with two particles on modes $(i,j)$, the scattering matrices $S$ will take elements from the $i$-th and $j$-th columns of $\tfrac{1}{\sqrt{2^p}}H(2^p)$. Let us put these two columns one next to the other, to form the $n \times 2$ matrix $A$. For a given output state $(i',j')$, the scattering matrix $S$ will be a submatrix of $A$ formed by its $i'$-th and $j'$-th rows. 

Note that two different columns of an Hadamard matrix have half of the elements with opposite sign and half of the elements with the same sign. Half of the rows of $A$ will be $\left[ 1,1 \right]$ or $\left[-1,-1\right]$; the other half will be $\left[-1,1\right]$ or $\left[1,-1\right]$. Let's now perform the following operations. First, we multiply the $\left[-1,-1\right]$ and $\left[-1,1\right]$ rows by -1. This will change the sign of the permanent of the scattering matrix that should include such rows, but it has no influence if the permanent vanishes. At this point we will have half of the rows equal to $\left[ 1,1 \right]$ and half equal to $\left[ 1, -1 \right]$. Second, we reorder the rows alternating $\left[ 1,1 \right]$ to $\left[ 1, -1 \right]$. This is equivalent to relabelling the outputs, which does not affect the number of suppressed output states. At this point the matrix $A$ will be just the same as if $(i,j) = (1,2)$, which is the case of Prop.~\ref{twoPhotonFirstTwoRows} and Prop.~\ref{twoFermionsFirstTwoRows}. Hence, every input combination has the same fraction of suppressed output states of the input $(i,j) = (1,2)$, discussed in Corol.~\ref{twoPhotonFraction} and Corol.~\ref{twoFermionsFraction} for boson and fermions respectively.\qed

\section{Multiple bosons} \label{sec:multiPhot}

The aim of this section is to demonstrate a suppression law for the case of $n=2^q$ indistinguishable bosons. This will be obtained in Prop.~\ref{suppressionLaw} at the end of the section. However, that result is based on other propositions which will be proved before. The first one (Prop.~\ref{largerMatrices}) allows to restrict the study, in certain conditions, from the case of $n$ particles in $m$ modes to the case of $n$ particles in $n$ modes. Props.~\ref{permanentSubmatrices}, \ref{nonVanishingPermanent}, on the contrary, regard mathematical properties (in particular, the value of the permanent) of certain $-1,+1$ matrices.

\begin{prop} \label{largerMatrices}
Be  $U = \tfrac{1}{\sqrt{2^p}}H(2^p)$ with $p = k + q$ a linear transformation over $m=2^p$ modes, and $\vec{h}=\left(1+n\cdot c,\ldots,n+n \cdot c \right)$, where $0 \leq c \leq (2^k - 1)$, an input state of $n=2^q$ particles. The output state $\vec{g} = \left(g_1, \ldots, g_n\right)$ is suppressed if and only if the output state $\vec{g}' = \left(g'_1, \ldots, g'_n \right)$ with $g'_i = \left[  (g_i-1) \mod n \right] + 1 $ is suppressed for the transformation $U' = \tfrac{1}{\sqrt{2^q}}H(2^q)$ with $n$ particles entering one per each mode.
\end{prop}

For input states of the kind $\vec{h}=\left(1,\ldots,n\right)$ (i.e., one particle per each of the first $n$ modes), the scattering matrices will be submatrices of the first $n$ columns of $U$. From the construction of $H(2^p) = H(2^{k + q}) = H(n \cdot 2^k)$ with \eqref{eq:sylvester}, it is clear that these first $n$ columns are just repetitions of $H(n)$.  For an output state with $n$ particles distributed on the modes $g_1 \ldots g_n$, the $i$-th row of the scattering matrix $S$ will be extracted from the $g_i$-th row of $U$. Since such rows repeat identically every $n$ rows (for that regards the first $n$ columns), the scattering matrix is the same  for all states $\vec{g} = \left(g_1, \ldots, g_n \right)$ having the same $g_1 \mod n, \ldots, g_n \mod n$. 
We can look for the smallest mode numbers giving this condition, which are $g'_i = \left[(g_i-1) \mod n \right] + 1$. In that case the scattering matrix is the one we would expect for the output state $\vec{g'} = \left(g'_1, \ldots, g'_n \right)$ defined as above, when entering with $n$ particles (one per each mode) in an interferometer implementing $U' = \tfrac{1}{\sqrt{2^q}}H(2^q)$.

Let's consider now the more general case $\vec{h}=\left(1+n\cdot c,\ldots,n+n \cdot c\right)$ with $0 \leq c \leq 2^{k-1}$. Again from the construction in \eqref{eq:sylvester} it can be observed that such columns will be repetitions of $\pm H(n)$ (with a succession of signs + and - that depends on $c$). If we properly change the signs of the rows (operation that is equivalent to add a $\pi$ phase term to certain outputs, which does not influence the probability modulus) these columns can be made identical to those of the case $\vec{h}=\left(1,\ldots,n\right)$, discussed above. Hence, the output distribution is the same\qed.

This result hold for both bosons and fermions because no hypotheses on the particles statistics have been adopted. In addition it can be exploited for a more precise generalisation of the results of Props.~\ref{twoPhotonFirstTwoRows} and \ref{twoFermionsFirstTwoRows} to a wider range of input states.

\begin{figure*}
\includegraphics{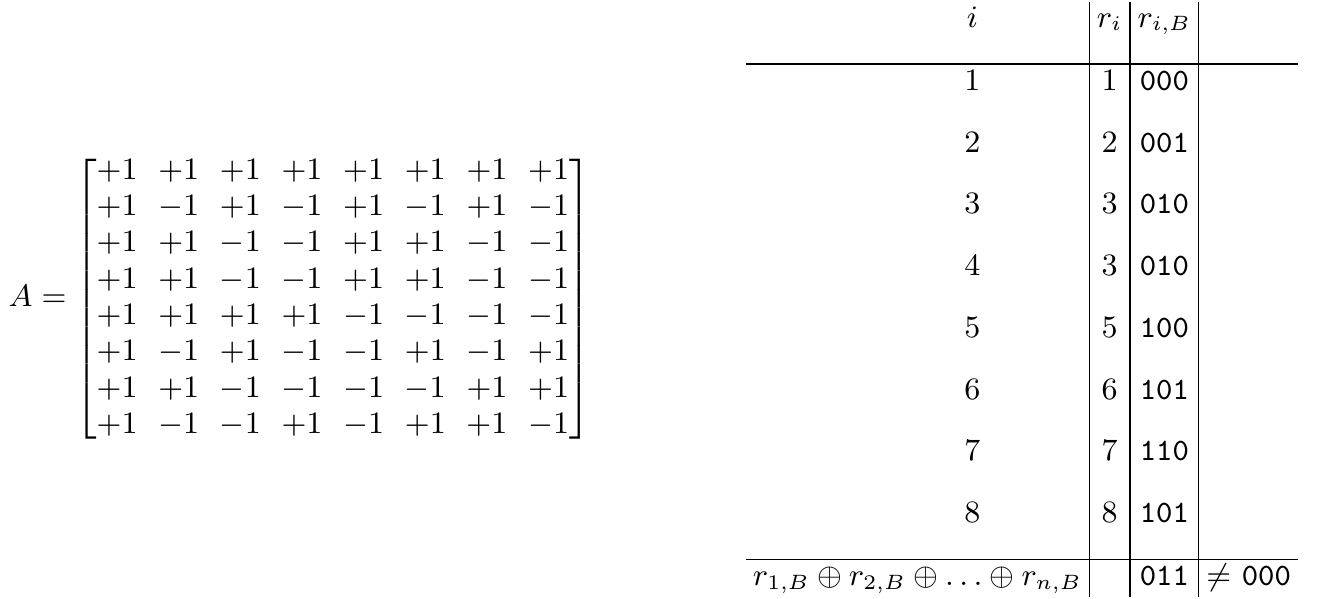}
\caption{Example of application of the criterion of  Prop.~\ref{permanentSubmatrices}, for a matrix $A$ built of the rows $\lbrace1,2,3,3,5,6,7,8\rbrace$ of $H(8)$. The table on the right summarise the application of the criterion: for each row $i$ of the scattering matrix, which is the $r_i$ line of $H(8)$, the binary expression $r_{i,B}$ is reported. The latter is actually the binary conversion of $r_i -1$, since the binary enumeration of the rows must start from 0. The last line of the table reports the bitwise sum which, being not equal to zero in this case, indicates that $\perm A = 0$.}
\label{fig:exampleCriterion}
\end{figure*}

\begin{prop} \label{permanentSubmatrices}
Be $A$ an $m \times m$ matrix, with $m=2^p$, built by taking the rows $\lbrace r_1, r_2, ..., r_m \rbrace$ from $H(m)$ (namely, the $i$-th row of $A$ is the $r_i$-th row of $H(m)$ and rows may be repeated). If $r_{1,B} \oplus r_{2,B} \oplus \ldots \oplus r_{m,B} \neq 0$, then $\perm A = 0$, being $r_{i,B}$ the binary representation of the row number, starting the count from 0, and $\oplus$ the bitwise sum (XOR operation).
\end{prop}

The condition $r_{1,B} \oplus r_{2,B} \oplus \ldots \oplus r_{m,B} \neq 0$ means that for at least one $k$, an odd number of $r_{i,B}$ has the same $k$-th bit. In other words, for at least one $k$, the $k$-th bit of the binary representations of the $r_i$ is $1$ for an odd number of rows and is $0$ for a (possibly different) odd number of rows. 

Consider now an arbitrary permutation $\sigma$ in the permanent expression $\perm  A = \sum_\sigma \prod^n_{i=1} a_{i,\sigma(i)}$, which is actually a set $\lbrace \sigma(i) \rbrace$ containing the numbers from 1 to $n$ in a certain order; further, be $\sigma'$ another permutation, obtained from $\sigma$ by changing the $k$-th bit in all its components $\sigma(i)$ (written in their binary representation). 
Let's analyse the effect of this bit flip. First, one should recall that, from its definition,
\begin{equation}
a_{i,j} = \left( H(m) \right)_{r_i,j}
\end{equation}
with $| a_{i,j} | = 1 \quad \forall \, i,j$. Depending on the value of the $k$-th bit of $r_{i,B}$, one has from \eqref{eq:sylvesterElement}:
\begin{align}
\left. r_{i,B} \right|_k = 1 \Rightarrow a_{i,\sigma(i)} = \left( H(m) \right)_{r_i,\sigma(i)} &= \notag \\
= -\left( H(m) \right)_{r_i,\sigma'(i)} &= -a_{i,\sigma'(i)} \\
\left. r_{i,B} \right|_k = 0 \Rightarrow a_{i,\sigma(i)} = \left( H(m) \right)_{r_i,\sigma(i)} &= \notag \\ 
= \left( H(m) \right)_{r_i,\sigma'(i)} &=a_{i,\sigma'(i)}
\end{align}

If, as in the case of the hypotheses, an odd number of $r_{i,B}$ has the $k$-th equal to 1, in the product $\prod^n_{i=1} a_{i,\sigma'(i)}$ an odd number of factors change their sign with respect to  $\prod^n_{i=1} a_{i,\sigma(i)}$, giving:
\begin{equation} \label{eq:ProdSignChange}
\prod^n_{i=1} a_{i,\sigma(i)} = - \prod^n_{i=1} a_{i,\sigma'(i)}
\end{equation}
This means that for each permutation $\sigma$ there exist another one $\sigma'$, biunivocally associated to $\sigma$, for which \eqref{eq:ProdSignChange} holds. Hence, in the sum over all the $\sigma$ of \eqref{eq:permDef}, half of the addends will have sign -1 and the other half +1, which implies $\perm A = 0$.\qed

An example of application of this criterion is given in \figref{fig:exampleCriterion}.

\begin{corol} \label{changeTheLine}
Be $A$ an $m \times m$ matrix, with $m = 2^p > 2$, built by taking the rows $\lbrace r_1, r_2, ..., r_m \rbrace$ from $H(m)$ (namely, the $i$-th row of $A$ is the $r_i$-th row of $H(m)$ and rows may be repeated), and $r_{1,B} \oplus r_{2,B} \oplus \ldots \oplus r_{m,B} = 0$. Build the matrix $B$ such that all the rows are the same as those of $A$ except the $i$-th, such $i$-th row being another arbitrary $r'_i$-th row of $H(m)$, with $r'_i \neq r_i$. Then, $B$ satisfies Prop.~ \ref{permanentSubmatrices}. 
\end{corol}

The condition $r_{1,B} \oplus r_{2,B} \oplus \ldots \oplus r_{m,B} = 0$ means that in the set of the binary representations $\lbrace r_{1,B}, r_{2,B}, ..., r_{m,B} \rbrace$ each bit recurs an even number of times with value 0 and an even number of times with value 1. The matrix $B$ is built by removing from $A$ its $i$-th row (which was the $r_i$-th of $H(m)$) and by replacing it with another arbitrary $r'_i$-th row of $H(m)$. The binary representation  $r'_{i,B}$ is different from $r_{i,B}$ for at least one bit, say, the $k$-th bit. Note that in the set $\lbrace r_{1,B}, r_{2,B}, ..., r_{m,B} \rbrace$ such $k$-th bit had the value 0 for an even number of times, and the value 1 for an even number of times: now that we have changed $r_i$ with $r'_i$ the $k$-th bit will have the value 0 for an odd number of times and the value 1 for (another) odd number times. This implies $r_{1,B} \oplus r_{2,B} \oplus \ldots \oplus r'_{i,B} \oplus \ldots \oplus r_{n,B} \neq 0$  and the hypotheses for Prop.~\ref{permanentSubmatrices} are verified.\qed

\begin{prop} \label{nonVanishingPermanent}
Be $A$ an $m \times m$, with $m = 2^p > 2$,  built by taking the rows $\lbrace r_1, r_2, ..., r_n \rbrace$ from $H(m)$ (namely, the $i$-th row of $A$ is the $r_i$-th row of $H(m)$ and rows may be repeated). If $r_{1,B} \oplus r_{2,B} \oplus \ldots \oplus r_{n,B} = 0$, then $\perm  A \neq 0$.
\end{prop}

Take the Laplace expansion of the permanent along an arbitrary $i$-th row:
\begin{equation}
\perm  A = \sum_j a_{i,j} \cdot \perm  M_{i,j} \label{eq:permanentLaplace}
\end{equation}
where $a_{i,j}$ is an element of $A$ and $M_{i,j}$ is the $i,j$ minor of $A$. In other terms, \eqref{eq:permanentLaplace} can be read as a dot product
\begin{equation}
\perm  A = \vec{a}_i \cdot \vec{c} \label{eq:permanentDotProduct}
\end{equation}
where $\vec{a}_i$ is the $n$-element vector given by the $i$-th row of $A$ and $\vec{c}$ is the vector with elements $c_j = \perm  M_{i,j}$.

Consider now $m$ different matrices $A_l$, built by replacing the $i$-th row of $A$ with the $l$-th row of $H(m)$. For $l = r_i$ one has $A_l = A_{r_i} = A$, while all the other $A_l$ will differ from $A$ by one row. When one calculates $\perm  A_l$ according to \eqref{eq:permanentLaplace}, the minors $M_{i,j}$ are always the same for every $A_l$, because just the $i$-th row is changing. Hence, in \eqref{eq:permanentDotProduct} the vector $\vec{c}$ is always the same for every $A_l$. The permanents of the different $A_l$ can be interpreted as the projection of such $\vec{c}$ onto different vectors, given by the $l$-th row of $H(m)$.

It is important to note that $\vec{c}$ is a non-zero vector. The elements of this vector are permanents of matrices $M_{i,j}$, which are squared (+1,-1) matrices of order $m-1$ and it has been shown \cite{simion1983opm} that if $m=2^p$, then no matrices of order $m-1$ exist with vanishing permanent.

The rows of $H(m)$ form a complete (orthogonal) basis of $\mathbb{R}^n$: a non-zero vector $\vec{c}$ has at least one non-zero projection on one of the vectors of the basis. We have already shown (Prop.~\ref{changeTheLine}) that, for $A_l$ defined as above with $l \neq r_i$, one has $\perm  A_l = 0$, i.e. the projection of $\vec{c}$ on all the rows of $H(m)$, except the $r_i$-th, is vanishing. It follows that the projection on the $r_i$-th row must be non-zero: this implies $\perm  A_{r_i} = \perm  A \neq 0$.\qed


\begin{table*}
\centering {\bf BOSONS} \smallskip
\begin{ruledtabular}
\begin{tabular}{*{8}{c}}
 \multirow{2}{*}{\quad$\boldsymbol{n}$\quad} & \multicolumn{6}{c}{$\boldsymbol{m}$}& \multirow{2}{*}{$\boldsymbol{\frac{n-1}{n}}$} \\[0.3ex]
 & \textbf{2} & \textbf{4} & \textbf{8} & \textbf{16} & \textbf{32} & \textbf{64} &  \\[0.3ex] \hline
\textbf{2} & $\tfrac{1}{3} \simeq 0.33$ & $\tfrac{4}{10} = 0.4 $ & $\tfrac{16}{36} \simeq 0.44$  & $\tfrac{64}{136} \simeq 0.47$ & $\tfrac{256}{528} \simeq 0.48$ & $\tfrac{1,024}{2,080} \simeq 0.49$ & 0.5 \\ [0.3ex]
\textbf{4} & & $\tfrac{24}{35} \simeq 0.69 $ & $\tfrac{240}{330} \simeq 0.73$ & $\tfrac{2,880}{3,876}\simeq 0.74$ & $\tfrac{39,168}{52,360}\simeq0.75$ & $\tfrac{574,464}{766,480}\simeq0.75$ & 0.75 \\ [0.3ex]
\textbf{8} & & & $\tfrac{5,600}{6,435}\simeq0.870$ & $\tfrac{428,736}{490,314}\simeq0.874$ & $\tfrac{53,829,888}{61,523,748}\simeq0.875$ & $\tfrac{9,309,189,120}{10,639,125,640}\simeq0.875$& 0.875 \\[0.3ex]
\end{tabular}
\end{ruledtabular}
\caption{Fraction of suppressed states over the possible output states, when injecting a $m$-modes Sylvester interferometer with $n$ photons in the first $n$ inputs. The number of suppressed states has been calculated by checking the criterion of Prop.~\ref{suppressionLaw} for each possible output state. In the last column the estimation from the formula \eqref{eq:fractionSuppressed} is given for comparison.} \label{table:suppressed}
\end{table*}
\begin{table*}
\centering {\bf FERMIONS} \smallskip
\begin{ruledtabular}
\begin{tabular}{*{8}{c}}
 \multirow{2}{*}{\quad$\boldsymbol{n}$\quad} & \multicolumn{6}{c}{$\boldsymbol{m}$}& \multirow{2}{*}{$\boldsymbol{1-\frac{n!}{n^n}}$} \\[0.3ex]
 & \textbf{2} & \textbf{4} & \textbf{8} & \textbf{16} & \textbf{32} & \textbf{64} &  \\[0.3ex] \hline
\textbf{2} & $\tfrac{2}{3} \simeq 0.67$ & $\tfrac{6}{10} = 0.6 $ & $\tfrac{20}{36} \simeq 0.56$  & $\tfrac{72}{136} \simeq 0.53$ & $\tfrac{272}{528} \simeq 0.52$ & $\tfrac{1,056}{2,080} \simeq 0.51$ & 0.5 \\ [0.3ex]
\textbf{4} & & $\tfrac{34}{35} \simeq 0.97 $ & $\tfrac{314}{330} \simeq 0.95$ & $\tfrac{3,620}{3,876}\simeq 0.93$ & $\tfrac{48,264}{52,360}\simeq0.92$ & $\tfrac{700,944}{766,480}\simeq0.91$ & 0.91 \\ [0.3ex]
\textbf{8} & & & $\tfrac{6,434}{6,435}>0.999$ & $\tfrac{490,058}{490,314}>0.999$ & $\tfrac{61,458,212}{61,523,748}\simeq0.999$ & $\tfrac{10,622,348,424}{10,639,125,640}\simeq0.998$& 0.998 \\[0.3ex]
\end{tabular}
\end{ruledtabular}
\caption{Fraction of suppressed states over the possible output states, when injecting a $m$-modes Sylvester interferometer with $n$ fermions in the first $n$ inputs. In the last column the asymptotic value $1-\frac{n!}{n^n}$ is given for comparison.}
\label{table:suppressedFermions}
\end{table*}

\begin{prop} \label{suppressionLaw}
Consider a unitary transformation $m=2^p$ modes $U_m = \tfrac{1}{\sqrt{2^p}} H(2^p)$ with $p = k + q$, and an input state with $n=2^q$ bosons $\vec{h}=\left(1+n\cdot c,\ldots,n+n \cdot c \right)$, where $0\leq c \leq (2^k - 1)$. The output state $\vec{g}=\left( g_1, g_2, \ldots, g_n \right)$ is suppressed if and only if $g_{1,B} \oplus g_{2,B} \oplus \ldots \oplus g_{n,B} \neq 0$, being $g_{i,B}$ the binary representation of $g_i-1$ (i.e. the binary representation of the output mode number, starting the count from 0) truncated to the $q$ least significant bits, and $\oplus$ the bitwise sum (XOR operation).
\end{prop}

As a first thing we address the case $m=n$, with $\vec{h}=\left(1, \ldots, n \right)$. Here the result comes directly from considering that the probability of an output configuration $\vec{g}=\left( g_1, g_2, \ldots, g_n \right)$ is proportional to $\perm S$ where $S$ is a matrix whose $i$-th row is the $g_i$-th row of $H(n)$. Because of Prop.~\ref{permanentSubmatrices} and \ref{nonVanishingPermanent} such permanent is vanishing if and only if $g_{1,B} \oplus g_{2,B} \oplus \ldots \oplus g_{n,B} \neq 0$ (where $g_{i,B}$ is the full binary expression of $g_i-1$, composed of $q$ bits) thus giving in this case the suppression of the corresponding output configuration.

By exploiting Prop.~\ref{largerMatrices}, this result can now be extended to the case $m=2^q>2^p=n$ and input states of the kind $\vec{h}=\left(1+n\cdot c,\ldots,n+n \cdot c \right)$. In particular, the condition of Prop.~\ref{largerMatrices} of considering the mode numbers \textit{modulo} $n$, implies that a criterion of the kind $g_{1,B} \oplus g_{2,B} \oplus \ldots \oplus g_{n,B} \neq 0$ can be applied if $g_{i,B}$ is the binary representation of the mode index, truncated to the $q$ least significant bits.\qed \medskip

To evaluate the fraction of output combinations that is suppressed we need to consider the set of all possible output states $\lbrace g_1, g_2, \ldots, g_n \rbrace$ and estimate when their binary expressions $\lbrace g_{1,B}, g_{2,B}, \ldots, g_{n,B} \rbrace$, truncated to the $q$ least significant bits (for the arguments discussed above), satisfies 
$G = g_{1,B} \oplus g_{2,B} \oplus \ldots \oplus g_{n,B} \neq 0$.
Adopting an approach similar to that of Ref.~\cite{tichy2010ztl}, we assume that in such a set, a certain $k$-th bit of the binary expression $g_{i,B}$ (consisting of $q$ bits) can take the values 0 or 1 with equal probability, independently from the values of the other bits. In other words, we assume that in each subset of states with a certain bit combination (for the bits other than the $k$-th), the number of output states for which the $k$-th bit is 0 is equal to the number of states for which that bit is 1.

Let's now consider the possible values of the binary expression $G$, starting from its first bit. That bit is the result of the $\oplus$ operation on $n$ bits (the first bit of each $g_{i,B}$). If we consider the full set of possible outputs, such $n$ bits will be 0 or 1 the same number of times. Thus, also the first bit of $G$ will be 0 for half of the possible output states and 1 for the other half. Those states for which the first bit of $G$ is 1 already satisfy $G \neq 0$, so they are suppressed. For the other ones, they may be suppressed if the $\oplus$ operation on other bits give 1. One then considers the second bit and with analogous arguments notes that it will be 0 for the half of the output states and 1 for the other half. One continues with the same procedure up to the $q$-th bit.
Hence, the overall fraction $\frac{N^\mathrm{bos}_\mathrm{supp}}{N_\mathrm{states}}$ of suppressed states will be $\frac{1}{2}$ (fraction of states which has the first bit of $R$ equal to 1) summed to $\frac{1}{2} \cdot \frac{1}{2}$ (fraction of states which has the first bit of $R$ equal to 0 and the second equal to 1) summed to $\frac{1}{2} \cdot \frac{1}{2} \cdot \frac{1}{2}$ (fraction of states which has the first and second bit equal to 0 and the third equal to 1) and so on. This gives:
\begin{equation} \label{eq:fractionSuppressed}
\frac{N^\mathrm{bos}_\mathrm{supp}}{N_\mathrm{states}} \sim  \sum^k_{x=1}  \frac{1}{2^x} = \frac{2^k-1}{2^k} = \frac{n-1}{n}
\end{equation}

Table~\ref{table:suppressed} reports the fraction of suppressed bosonic states for $n \leq 8$ and $m \leq 64$, compared with the result of \eqref{eq:fractionSuppressed}. The latter expression approximates better the actual value for increasing $n$ or $m$.

\section{Multiple fermions} \label{sec:fermions}

The case of fermions is less significant, with respect to bosons, from a computational point of view; in fact the probability amplitude of the output configurations are proportional to the determinant of the scattering matrix (see Eq.~\eqref{eq:determinant}), which differently from the permanent can be calculated efficiently. However, investigating the suppression laws arising for this kind of particles enable a better understanding of the effects of statistics in multi-particle interference.

\begin{prop} \label{lawFermion}
Consider a unitary transformation over $m=2^p$ modes $U_m = \tfrac{1}{\sqrt{2^p}}H(2^p)$ with $p = k + q$, and an input state with $n=2^q$ fermions $\vec{h}=\left(1+n\cdot c,\ldots,n+n \cdot c \right)$, where $0 \leq c \leq (2^k - 1)$. The output state $\vec{g} = \left(g_1, \ldots, g_n\right)$ is suppressed if and only if $g_i \mod n = i \mod n \quad \forall i \in \left[1,n\right] $.
\end{prop}

Let's consider, to begin, the case $n = m$, i.e. $n$ fermions entering a $n$-mode interferometer one per each port. 
The only possible output state allowed from the Pauli principle is $\vec{g} = \left(1, \ldots, n \right)$, namely the state having one particle per mode, which can be written also as $g_i = i \quad \forall i \in \left[1,n\right]$. Such condition is easily extended to the case of a more generic input state $\vec{h}=\left(1+n\cdot c,\ldots,n+n \cdot c \right)$ of $n = 2^q$ fermions entering a $m= n \cdot 2^k$ interferometer through Prop.~\ref{largerMatrices}, becoming $g_i \mod n = i \mod n \quad \forall i \in \left[1,n\right]$.\qed
\medskip

The number of allowed/suppressed output states can be evaluated considering that, in an interferometer with $m = n \cdot 2^k$ modes, the condition $g_i \mod n = i \mod n$ can be satisfied for $2^k$ different values of $g_i$. Hence, the number of allowed output states is the number of sequences of $n$ numbers, each with $2^k$ possible values, i.e. $2^{k \cdot n} = \left(\frac{m}{n}\right)^n$.

The possible output states of $n$ particles on $m$ modes are $N_\mathrm{states} =\binom{m+n-1}{n} = \frac{(m+n-1)!}{(m-1)!\,n!}$ (combinations with repetitions). The fraction of allowed states, for large $m$, thus tends to\footnote{Note that, for large $a$, one has
\begin{align*} 
\frac{a!}{(a+b)!} = \frac{1}{(a+b)(a+b-1)\ldots(a+1)} &=\\
= \frac{1}{a^b(1+\frac{b}{a})(1+\frac{b-1}{a})\ldots(1+\frac{1}{a})} &\asymp \frac{1}{a^b}
\end{align*}.
}:
\begin{equation}
\left(\frac{m}{n}\right)^n \cdot \frac{(m-1)!\,n!}{(m+n-1)!} \asymp \left(\frac{m}{m-1}\right)^n  \cdot \frac{n!}{n^n} \asymp \frac{n!}{n^n}
\end{equation}
and the fraction of suppressed states is asymptotically equal to:
\begin{equation}
\frac{N^\mathrm{fer}_\mathrm{supp}}{N_\mathrm{states}} = 1 - \frac{N^\mathrm{fer}_\mathrm{allowed}}{N_\mathrm{states}}  \asymp 1-\frac{n!}{n^n}
\label{eq:ferSuppressedFraction}
\end{equation} 

Table~\ref{table:suppressedFermions} reports the fraction of suppressed fermionic states for $n \leq 8$ and $m \leq 64$, calculated over the all the possible $n$-particles states, together with the asymptotic estimation with \eqref{eq:ferSuppressedFraction}.

\section{Discussion} \label{sec:disc}

\begin{figure}
\centering
\includegraphics{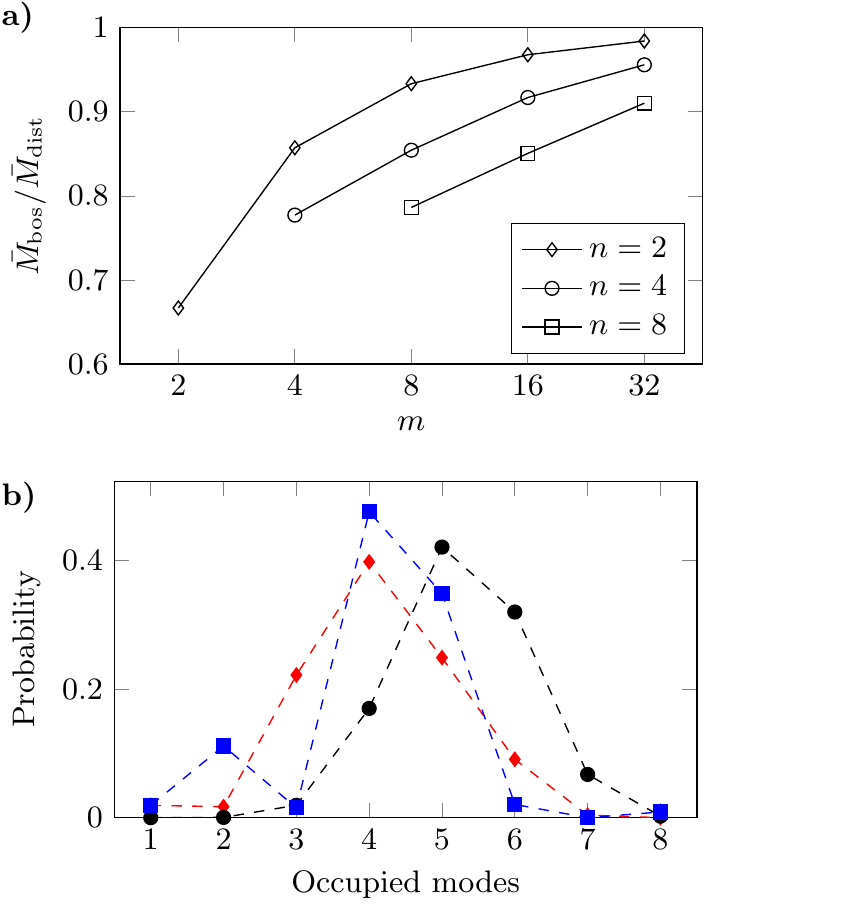}
\caption{a) Ratio between the average number of occupied modes for bosons and for distinguishable particles, for $n=2,4,8$ particles entering an $m$-mode Sylvester interferometer, one per each of the first $n$ modes. b) Probability distribution of detecting an output state with a certain number of occupied modes, when entering with $n=8$ indistinguishable bosons (one per each input mode) in a $m = 8$ Sylvester interferometer (blue squares) or Fourier interferometer (red diamonds). Probability for distinguishable particles (black circles) in analogous interferometers is also reported.}
\label{fig:occupNumber}
\end{figure}

The predictions of the suppressions laws proved above can be compared with the general trends that usually distinguish the particles behaviour, depending on their statistics. While for fermions the compatibility of Prop.~\ref{lawFermion} with the Pauli principle is intrinsic in its same proof, the behaviour of bosons, which would be expected to show an enhanced tendency to bunch together, is more peculiar. Actually, similarly to Ref.~\cite{tichy2010ztl}, we observe that many bunching events are not enhanced but included in the suppression conditions of Prop.~\ref{suppressionLaw}. The probability of full-bunching events (events with all $n$ bosonic particles on the same output mode, over the $m$ possible modes) can be calculated easily: the scattering matrix is composed of identical rows with half $+1/\sqrt{m}$ and half $-1/\sqrt{m}$ elements; by inverting the sign of the columns with negative elements (operation which does not affect the permanent \footnote{In fact, each sign inversion of a column inverts the sign of the permanent, but this is performed an even number of times.}), we obtain a matrix of all $+1/\sqrt{m}$, whose permanent equals $n!/m^{m/2}$. Thus the probability of this event is $\left(\tfrac{n!}{m^{m/2}\sqrt{n!}}\right)^2 = n!/m^m$ (from squaring Eq.~\ref{eq:permanent}) which is an enhancement of $n!$ with respect to the probability of such event for distinguishable particles. This enhancement factor for full bunching events is indeed a general law for all unitary processes \cite{spagnolo2013grb}. With regard to anti-bunching events (particles in all different ports), it is not difficult to observe that for $n = m > 2$  they are instead never suppressed for bosons: the scattering matrix for such an event would be a full Sylvester matrix, whose permanent is proved \footnote{Incidentally, we note that this is also a partial answer to an open problem regarding permanents \cite{cheon2005ums}, i.e., the question whether the permanent of a Hadamard matrix could in general vanish.} to be non-vanishing by observing that for $\lbrace r_1, r_2, \ldots, r_n \rbrace=\lbrace 1,2,\ldots,n \rbrace$ one has $r_{1,B} \oplus r_{2,B} \oplus \ldots \oplus r_{n,B} = 0$ and by exploiting Prop.~\ref{nonVanishingPermanent}. This also marks a difference with respect to Bell multiports, where such events are always suppressed for an even number of bosons \cite{lim2005gho}.

An overall figure that may quantify the bunching behaviour \cite{tichy2010ztl} is the average number $\bar{M}$ of occupied modes at the output. The bosonic bunching tendency should reduce the number of occupied modes with respect to the case of distinguishable particles. Figure~\ref{fig:occupNumber}a reports the ratio between the average number of occupied modes in the case of indistinguishable bosons and that in the case of distinguishable particles, for $n=2,4,8$ particles injected in the first modes of Sylvester interferometers with up to 32 modes. 
This ratio is always smaller than unity, confirming the bunching behaviour. However, while from the previous discussion we know that the fraction of suppressed states is practically constant with increasing $m$, this ratio looks to approach one in the same limit: the larger number of available modes makes the particles more likely to exit on different ports, approaching the classical probability on this aspect. Thus, the suppression law seems to be a stronger non-classical signature than the bunching behaviour itself, which is less evident with large $m$.

The detailed probability of having a certain number of occupied modes is reported in Figure~\ref{fig:occupNumber}b for the case $n=m=8$, with the comparison of the distribution in the case of a Fourier interferometer. The three distributions are different in shape and it is evident the shift towards a smaller number of occupied modes for the two non-classical distributions. In this particular case the average number of occupied ports is $\simeq 4.1$ for both Sylvester and Fourier interferometers in case of identical bosons and is $\simeq 5.3$ in case of distinguishable particles.

As a further analysis, the asymptotic fraction of allowed fermionic states can be compared with the asymptotic fraction of allowed bosonic states, to evaluate the strength of the suppression law in the two cases. For large $m$, one reads:
\begin{equation} \label{eq:comparison}
\frac{N^\mathrm{fer}_\mathrm{allowed}}{N_\mathrm{states}} \asymp \frac{n!}{n^n} = \frac{n}{n}\cdot\frac{n-1}{n}\cdot\ldots\cdot\frac{1}{n} < \frac{1}{n} \asymp \frac{N^\mathrm{bos}_\mathrm{allowed}}{N_\mathrm{states}}
\end{equation}
Interestingly, the suppression law seems to act more severely for fermions, for large $m$. Note that the calculation in \eqref{eq:comparison} actually includes, for fermions, all the possible multi-particle states, even those already forbidden from the Pauli principle itself. It can be observed, however, that $N^\mathrm{fer}_\mathrm{states} = \binom{m}{n} \asymp \binom{m+n-1}{n} = N_\mathrm{states}$ for large $m$, where $N^\mathrm{fer}_\mathrm{states}$ are all the possible fermionic states, i.e. states with all different output ports. Thus the inequality \eqref{eq:comparison} holds asymptotically also considering (for fermions) only the fraction of events that were not already suppressed by the simple application of the Pauli principle.
It is worth reminding that, besides manifesting naturally for true fermions and bosons, the effects of the statistics can be simulated by proper entangled states \cite{omar2006qwo, matthews2013ofs}. Thus, the laws here developed for the two kinds of particles hold true for the corresponding entangled states. 

From a more applicative point of view, suppression laws such as the one presented here for Sylvester interferometers may be exploited to test the indistinguishability of $n$-photon sources\cite{tichy2010ztl}. Further, they may be used, in the context of boson sampling experiments, to simultaneously check the quality of the sources and of a possible reconfigurable device \cite{tichy2014sea}, that would perform the required unitary that expresses the suppression law. The limitation of the laws for Sylvester interferometers to specific values of $n$ may seem, at first glance, quite disadvantageous in a possible general case when it may be required to test a general $n$-photon source where $n$ may not be a power of 2. However, one may envisage that the full $n$-photon interference could be used as an overall check of the source quality, but an accurate troubleshooting of possible malfunctioning or imperfections requires different subsets of single photon sources to be tested separately. To this purpose one may configure a device to implement a block-diagonal matrix, having for each block a different $2^p$-mode (Sylvester) unitary to test separately different subsets of single-photons.

An interesting perspective to test at the same time the indistinguishability of different couples of photons, without the need of reconfiguring the device is given by Prop.~\ref{twoPhotonArbitraryInputs}, which is characteristic of Sylvester matrices and does not hold for Fourier ones. In fact, in the two photon case, whichever couple of inputs is excited, an identical fraction of outputs is suppressed. This may be particularly useful to check single-photon sources for scattershot boson sampling \cite{lund2014bsg}: there, several heralded single-photon sources are used, each coupled to a different input port, and multi-photon states (with one photon per port, but multiple ports excited) are generated randomly. Couples of photons on random input ports are generated efficiently in such a setup and by comparing the detected output events with the predictions of Prop.~\ref{twoPhotonArbitraryInputs} the indistinguishability of all the possible couples of sources may be conveniently tested.

\section{Conclusion} \label{sec:concl}

In conclusion, we have proved a necessary and sufficient criterion for the suppression of many output combinations when $n = 2^q$ particles are injected in certain inputs of a linear interferometer with $m=2^p$ modes implementing a Sylvester matrix. While both the bosonic and fermionic cases have been studied, the result is particularly significant for bosons, whose output distribution is hard to compute in the general case. Therefore, this suppression law may be exploited for the use of Sylvester multiports as benchmark devices for the indistinguishability of multiple single-photon sources or the assessment of the overall quality of reconfigurable interferometers.

This study has also shown that comprehensive laws that describe the output multi-photon distribution of multi-port interferometers on the basis of the symmetry of the implemented matrix are not limited to Fourier ones. Indeed, further investigations could pursue the definition of similar criteria for wider class of matrices, thus giving greater insight on the features of multi-particle interference.

\begin{acknowledgments}
The author acknowledges financial support from the ERC-Starting Grant 3D-QUEST (3D-Quantum Integrated Optical Simulation; grant agreement no. 307783): \texttt{http://www.3dquest.eu}.
\end{acknowledgments}

%

\end{document}